\documentclass[]{jetpl}
\twocolumn
\usepackage{epsf}

\lat
\newcommand{\diverg}{\mathop{\rm div}\nolimits}
\newcommand{\const}{\mathop{\rm const}\nolimits}
\renewcommand{\vec}{\mathbf}

\title{Decay of the monochromatic capillary wave.}

\rtitle{Decay of the monochromatic capillary\ldots}

\sodtitle{Decay of the monochromatic capillary wave}

\author{A.\,I.\,Dyachenko$^{+}$,
A.\,O.\,Korotkevich$^+$\/\thanks{e-mail: kao@landau.ac.ru}, V.\,E.\,Zakharov$^{+*}$}

\rauthor{A.\,I.\,Dyachenko, A.\,O.\,Korotkevich, V.\,E.\,Zakharov}

\sodauthor{Dyachenko, Korotkevich, Zakharov}

\address{$^+$L.D. Landau Institute for Theoretical Physics RAS,
119334 Moscow, Russia\\~\\
$^*$University of Arizona, Department of Mathematics, Tucson,  USA
}

\dates{27 March 2003}{*}

\abstract{It was demonstrated by direct numerical simulation that, in the case of
weakly nonlinear capillary waves, one can get resonant waves interaction on the discrete
grid when resonant conditions are never fulfilled exactly. The waves's decay pattern was obtained. The influence of the mismatch of resonant condition was studied as well.}


\begin{document}

\maketitle

Nonlinear waves on the surface of a fluid are one of the most well known and complex phenomena in
nature. Mature ocean waves and ripples on the surface of the tea in a pot, for example, can be
described by very similar equations. Both these phenomena are substantially nonlinear,
but the wave amplitude is usually significantly less than the wavelength. Under this condition, waves
are weakly nonlinear.

To describe processes of this kind, the weak turbulence theory was proposed \cite{Zakharov-DAN66},\cite{Zakharov-JAMTP67}.
It results in Kolmogorov spectra as an exact solution of the Hasselman-Zakharov kinetic equation \cite{Springer-92}.
Many experimental results are in great accordance with this theory. In the case of gravity surface waves, the first confirmation was obtained by Toba \cite{Toba}, and the most recent data by Hwang \cite{Hwang} were obtained as a result of lidar scanning of the ocean surface. Recent experiments with capillary waves on the surface of liquid hydrogen \cite{Kolmakov-Lett},\cite{Kolmakov} are also in good agreement with this theory.
On the other hand, some numerical calculations have been made to check the validity of the
weak turbulent theory \cite{Pushkarev-96},\cite{Dias-2001},\cite{Vasiliev-2002}.

In this Letter we study the one of the keystones of the weak turbulent theory,
the resonant interaction of weakly nonlinear waves.
The question under study is the following: 
\begin{itemize}
\item How does a discrete grid for wavenumbers in numerical
simulations affects the resonant interaction?
\item Can a nonlinear frequency shift broad resonant
manifold to make discreteness unimportant?
\end{itemize}
We study this problem for nonlinear capillary
waves on the surface of an infinite depth incompressible ideal fluid. Direct numerical
simulation can make the situation clear.

Let us consider the irrotational flow of an ideal incompressible fluid of infinite depth. For the sake
of simplicity, let us suppose fluid density $\rho = 1$. The velocity potential $\phi$ satisfies the Laplace equation
\begin{equation}
\label{Laplas}
\triangle \phi = 0
\end{equation}
in the fluid region bounded by
\begin{equation}
-\infty < z < \eta (\vec r), \;\;\;\; \vec r = (x,y),
\end{equation}
with the boundary conditions for the velocity potential
\begin{equation}
\label{Laplas_boundary}
\begin{array}{c}
\displaystyle
\left. 
\frac {\partial \eta}{\partial t}+
\frac {\partial \phi}{\partial x}\frac {\partial \eta}{\partial x} +
\frac {\partial \phi}{\partial y}\frac {\partial \eta}{\partial y}
= \frac{\partial \phi}{\partial z} \right|_{z= \eta},\\
\displaystyle
\left. \left ( 
\frac {\partial \phi}{\partial t}+
\frac{1}{2}( \nabla \phi )^2
\right ) \right |_{z= \eta} +\\
\displaystyle
+\sigma (\sqrt{1 + (\nabla \eta)^2} - 1)
= 0,
\end{array}
\end{equation}
on $z=\eta$, and
\begin{equation}
\phi_z|_{z= - \infty} = 0,
\end{equation}
on $z\to -\infty$.
Here $\eta = \eta (x,y,t)$ is the surface displacement.
In the case of capillary waves, the Hamiltonian has the form
$$
H = T + U,
$$
\begin{equation}
T = \frac{1}{2} \int d^2 r \int \limits_{-\infty}^{\eta} (\nabla \phi)^2 dz,
\end{equation}
\begin{equation}
U = \sigma \int (\sqrt{1 + (\nabla \eta)^2} - 1)d^2 r,
\end{equation}
where $\sigma$ -- is the surface tension coefficient.
In \cite{Zakharov-68}, it was shown that this system is Hamiltonian. The Hamiltonian
variables are the displacement of the surface $\eta(x,y,t)$ and velocity potential
on the surface of the fluid $\psi (x,y;t) = \phi (x,y,\eta(x,y;t);t)$. Hamiltonian equations are
\begin{equation}
\label{Hamiltonian_equations}
\frac{\partial \eta}{\partial t} = \frac{\delta H}{\delta \psi}, \;\;\;\;
\frac{\partial \psi}{\partial t} = - \frac{\delta H}{\delta \eta}.
\end{equation}
Using the weak nonlinearity assumption \cite{Springer-92} one can expand the Hamiltonian
in the powers of surface displacement
\begin{equation}
\label{Hamiltonian}
\begin{array}{l}
\displaystyle
H = \frac{1}{2}\int\left( \sigma |\nabla \eta|^2 + \psi \hat k  \psi \right) d^2 r + \\
\displaystyle
+ \frac{1}{2}\int\eta\left[ |\nabla \psi|^2 - (\hat k \psi)^2 \right] d^2 r.
\end{array}
\end{equation}
The third order is enough for three-wave interactions.
Here, $\hat k$ is the linear operator corresponding to multiplication of Fourier harmonics by
the modulus of the wavenumber $\vec k$. Using (\ref{Hamiltonian_equations}),
one can get the following system of dynamical equations:
\begin{equation}
\label{eta_psi_system}
\begin{array}{l}
\displaystyle
\dot \eta = \hat k  \psi - \diverg (\eta \nabla \psi) - \hat k  [\eta \hat k  \psi],\\
\displaystyle
\dot \psi = \sigma \triangle \eta - \frac{1}{2}\left[ (\nabla \psi)^2 - (\hat k \psi)^2 \right]
\end{array}
\end{equation}
The properties of $\hat k$-operator suggest exploiting the equations in Fourier space for
Fourier components of $\eta$ and $\psi$,
$$
\psi_{\vec k} = \frac{1}{2\pi} \int \psi_{\vec r} e^{i {\vec k} {\vec r}} d^2 r,\;\;
\eta_{\vec k} = \frac{1}{2\pi} \int \eta_{\vec r} e^{i {\vec k} {\vec r}} d^2 r.
$$
Let us introduce the canonical variables $a_{\vec k}$ as shown below
\begin{equation}
a_{\vec k} = \sqrt \frac{\omega_k}{2k} \eta_{\vec k} + i \sqrt \frac{k}{2\omega_k} \psi_{\vec k},
\end{equation}
where
\begin{equation}
\omega_k = \sqrt {\sigma k^3}.
\end{equation}
With these variables, the Hamiltonian  (\ref{Hamiltonian}) acquires the form
\begin{equation}
\begin{array}{l}
\displaystyle
H = \int \omega_k |a_{\vec k}|^2 d {\vec k} + \\
\displaystyle
+\frac{1}{6}\frac{1}{2\pi}\int E_{\vec k_1 \vec k_2}^{\vec k_0} 
(a_{\vec k_1}a_{\vec k_2}a_{\vec k_0} + a_{\vec k_1}^{*}a_{\vec k_2}^{*}a_{\vec k_0}^{*})\times\\
\displaystyle
\times\delta (\vec k_1 + \vec k_2 + \vec k_0) d {\vec k_1}d {\vec k_2}d {\vec k_0} +\\
\displaystyle
+\frac{1}{2}\frac{1}{2\pi}\int M_{\vec k_1 \vec k_2}^{\vec k_0} 
(a_{\vec k_1}a_{\vec k_2}a_{\vec k_0}^{*} + a_{\vec k_1}^{*}a_{\vec k_2}^{*}a_{\vec k_0})\times\\
\displaystyle
\times\delta (\vec k_1 + \vec k_2 - \vec k_0) d {\vec k_1}d {\vec k_2}d {\vec k_0}.
\end{array}
\end{equation}
Here,
\begin{equation}
\begin{array}{l}
\displaystyle
E_{\vec k_1 \vec k_2}^{\vec k_0} = V_{\vec k_1 \vec k_2}^{k_0} + V_{\vec k_0 \vec k_2}^{k_1} + V_{\vec k_0 \vec k_1}^{k_2},\\
\displaystyle
M_{\vec k_1 \vec k_2}^{\vec k_0} = V_{\vec k_1 \vec k_2}^{k_0} - V_{-\vec k_0 \vec k_2}^{k_1} - V_{-\vec k_0 \vec k_1}^{k_2},\\
\displaystyle
V_{\vec k_1 \vec k_2}^{k_0} = \sqrt {\frac{\omega_{k_1} \omega_{k_2} k_0}{8 k_1 k_2 \omega_{k0}}} L_{\vec k_1 \vec k_2},\\
\displaystyle
L_{\vec k_1 \vec k_2} = (\vec k_1 \vec k_2) + |k_1||k_2|.
\end{array}
\end{equation}
The dynamic equations in this variables can be easily obtained by variation of Hamiltonian
\begin{equation}
\begin{array}{l}
\displaystyle
\dot a_{\vec k} = -i \frac{\delta H}{\delta a_{\vec k}^{*}} = -i \omega_k a_{\vec k} - \\
\displaystyle
-\frac{i}{2}\frac{1}{2\pi}\int M_{\vec k_1 \vec k_2}^{\vec k} 
a_{\vec k_1}a_{\vec k_2}\delta (\vec k_1 + \vec k_2 - \vec k) d {\vec k_1}d {\vec k_2}-\\
\displaystyle
-\frac{i}{2\pi}\int M_{\vec k \vec k_2}^{\vec k_0} 
a_{\vec k_2}^{*}a_{\vec k_0}\delta (\vec k + \vec k_2 - \vec k_0) d {\vec k_2}d {\vec k_0}-\\
\displaystyle
-\frac{i}{2}\frac{1}{2\pi}\int E_{\vec k_1 \vec k_2}^{\vec k} 
a_{\vec k_1}^{*}a_{\vec k_2}^{*}\delta (\vec k_1 + \vec k_2 + \vec k) d {\vec k_1}d {\vec k_2}.
\end{array}
\end{equation}
Each term in this equation has its own clear physical meaning. The linear term gives a periodic evolution
of the initial wave. The first nonlinear term describes a merging of two waves $\vec k_1$ and $\vec k_2$
in $\vec k$.
The second describes a decay of the wave $\vec k_0$ to the waves $\vec k$ and $\vec k_2$.
And the last term corresponds to the second harmonic generation process. It is useful to eliminate the linear term with the substitution
\begin{equation}
a_{\vec k} = A_{\vec k} e^{i \omega_k t}.
\end{equation}
In this variables, the dynamical equations take the form
\begin{equation}
\label{A-equations_continuous}
\begin{array}{l}
\displaystyle
\dot A_{\vec k} = -\frac{i}{2}\frac{1}{2\pi}\int M_{\vec k_1 \vec k_2}^{\vec k} 
A_{\vec k_1}A_{\vec k_2}e^{i\Omega_{k_1 k_2}^{k} t}\times\\
\displaystyle
\times\delta (\vec k_1 + \vec k_2 - \vec k) d {\vec k_1}d {\vec k_2}-\\
\displaystyle
-\frac{i}{2\pi}\int M_{\vec k \vec k_2}^{\vec k_0} 
A_{\vec k_2}^{*}A_{\vec k_0}e^{-i\Omega_{k k_2}^{k_0} t}\times\\
\displaystyle
\times\delta (\vec k + \vec k_2 - \vec k_0) d {\vec k_2}d {\vec k_0},
\end{array}
\end{equation}
where
\begin{equation}
\Omega_{k_1 k_2}^{k_0} = \omega_{k_1} + \omega_{k_2} - \omega_{k_0}.
\end{equation}
Here we do not consider the harmonic generation term. The remaining terms give us
the following conditions of resonance
\begin{equation}
\label{resonant_conditions}
\Omega_{k_1 k_2}^{k} = \omega_{k_1} + \omega_{k_2} - \omega_{k} = 0,\;\;\;
\vec k_1 + \vec k_2 - \vec k = 0.
\end{equation}
All this theory is well known in the literature \cite{Springer-92}.

Now let us turn to the discrete grid.
Also, from this point we assume periodic boundary conditions in $x$ and $y$ with
lengths $L_x$ and $L_y$. One can easily obtain equations similar to (\ref{A-equations_continuous})
\begin{equation}
\begin{array}{l}
\displaystyle
\dot A_{\vec k} = -\frac{i}{2}\frac{2\pi}{L_x L_y}\sum\limits_{\vec k_1 \vec k_2} M_{\vec k_1 \vec k_2}^{\vec k} 
A_{\vec k_1}A_{\vec k_2}e^{i\Omega_{k_1 k_2}^{k} t}\times\\
\displaystyle
\times\triangle_{(\vec k_1 + \vec k_2), - \vec k} -\\
\displaystyle
-\frac{i2\pi}{L_x L_y}\sum\limits_{\vec k_2 \vec k_0} M_{\vec k \vec k_2}^{\vec k_0} 
A_{\vec k_2}^{*}A_{\vec k_0}e^{-i\Omega_{k k_2}^{k_0} t}\triangle_{(\vec k + \vec k_2), - \vec k_0},
\end{array}
\end{equation}
where $\triangle_{\vec k_1, \vec k_2}$ is the Kronecker delta -- the discrete analogue of the Dirac delta function.

Consider the decay of a monochromatic capillary wave $A_{\vec k_o}$ on two
waves
\begin{equation}
\label{two-waves_decay}
\begin{array}{l}
\displaystyle
\dot A_{\vec k_0} = -\frac{i}{2}\frac{2\pi}{L_x L_y} M_{\vec k_1 \vec k_2}^{\vec k_0} 
A_{\vec k_1}A_{\vec k_2}e^{i\Omega_{k_1 k_2}^{k_0} t},\\
\displaystyle
\dot A_{\vec k_1} = -i\frac{2\pi}{L_x L_y} M_{\vec k_1 \vec k_2}^{\vec k_0} 
A_{\vec k_2}^{*}A_{\vec k_0}e^{-i\Omega_{k_1 k_2}^{k_0} t},\\
\displaystyle
\dot A_{\vec k_2} = -i\frac{2\pi}{L_x L_y} M_{\vec k_1 \vec k_2}^{\vec k_0} 
A_{\vec k_1}^{*}A_{\vec k_0}e^{-i\Omega_{k_1 k_2}^{k_0} t}.
\end{array}
\end{equation}
Let $A_{\vec k_1}, A_{\vec k_2}$ be small ($\left|A_{\vec k_0}\right| \gg \max (|A_{\vec k_1}|, |A_{\vec k_2}|)$ at $t=0$).
In this case the equations can be linearized. The solution of linearized (\ref{two-waves_decay})
has the form ($A_{\vec k_0} \sim \const$)
\begin{equation}
\label{exp-growth}
A_{\vec k_{1,2}}(t) = A_{\vec k_{1,2}}(0) e^{\lambda t},
\end{equation}
where
\begin{equation}
\label{lambda}
\begin{array}{c}
\displaystyle
\lambda =-\frac{i}{2}\Omega_{k_1 k_2}^{k_0} + \\
\displaystyle
+\sqrt{\left|\frac{2\pi}{L_x L_y} M_{\vec k_1 \vec k_2}^{\vec k_0} A_{\vec k_0}\right|^2 - \left(\frac{1}{2}\Omega_{k_1 k_2}^{k_0}\right)^2}.
\end{array}
\end{equation}

In the case of a continuous media, resonant conditions (\ref{resonant_conditions}) can be satisfied exactly.
But on the grid, there is always a frequency mismatch $\Omega_{k_1 k_2}^{k_0} \ne 0$ although if the amplitude of the initial wave is high enough there are resonances even on a discrete grid.  But the width of this resonance is very important. 

System of equations (\ref{eta_psi_system}) can be solved numerically. 
This system is nonlocal in coordinate space due to the presence of the $\hat k $-operator. 
The origin of this operator gives us a hint to solve (\ref{eta_psi_system}) in wavenumbers space ($K$-space). 
In this case we can effectively use the fast Fourier transform algorithm. Omitting the details of this numerical 
scheme, we reproduce only the final results of calculations.

We have solved system of equations (\ref{eta_psi_system}) numerically in the periodic
domain $2\pi\times2\pi$ (the wave-numbers $k_x$ and $k_y$ are integer numbers in this case). The size of the grid was chosen as $512\times512$ points. We have also included damping for waves with large wave numbers. In $K$-space damping  terms for $\eta_{\vec k}$ and $\psi_{\vec k}$ 
respectively were the following: $\gamma_{\vec k} \eta_{\vec k}$ and $\gamma_{\vec k} \psi_{\vec k}$,
where $\gamma_{\vec k}$ was of the form
\begin{equation}
\begin{array}{c}
\displaystyle
\gamma_{\vec k} = 0, |{\vec k}| < \frac{1}{2} |{\vec k_{max}}|,\\
\displaystyle
\gamma_{\vec k} = - \gamma_0 (|{\vec k}| - |\frac{1}{2}{\vec k_{max}}|)^2, 
|{\vec k}| \ge \frac{1}{2} |{\vec k_{max}}|,
\end{array}
\end{equation}
here, $\gamma_0$ is some constant.

As an initial conditions we used one monochromatic wave of sufficiently large amplitude
with wave numbers $\vec k_0$ ($k_{0x} = 0, k_{0y} = 68$). Along with that there was a small random noise in all other harmonics.

\begin{figure}[hbt]
\centerline{\epsfxsize=8.5cm \epsfbox{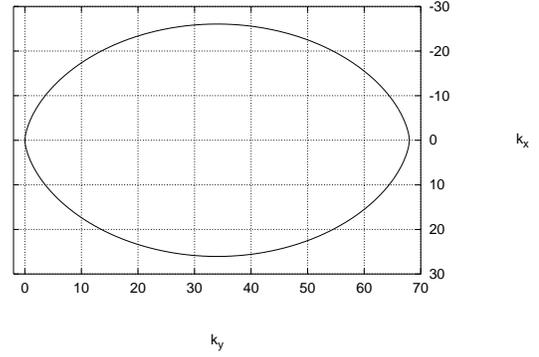}} 
\vspace{4mm}
\caption[]{\label{resonance_curve}Fig.1. The resonant manifold for $k_0 = 68$.}
\end{figure}

\begin{figure}[hbt]
\centerline{\epsfxsize=8.5cm \epsfbox{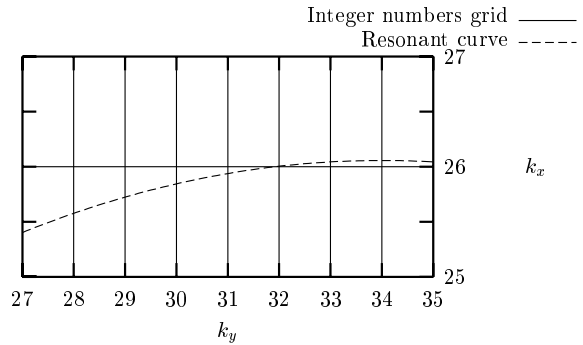}} 
\vspace{4mm}
\caption[]{\label{resonance_curve_local}Fig.2. Different mismatch is seen at different grid points.}
\end{figure}

Resonant manifold (\ref{resonant_conditions}) for decaying waves
\begin{equation}
\begin{array}{l}
\displaystyle
\vec k_0 = \left(
\begin{array}{c}
0\\
k_0
\end{array}
\right),\\
\displaystyle
\vec k_1 = \left(
\begin{array}{c}
-k_x\\
k_0 - k_y
\end{array}
\right),\;
\vec k_2 = \left(
\begin{array}{c}
k_x\\
k_0 + k_y
\end{array}
\right).
\end{array}
\end{equation}
is given at Fig.\ref{resonance_curve}.
Since the wave numbers are integers, the resonant curve never coincides with grid points
exactly. A detailed picture is given in Fig.\ref{resonance_curve_local}. It is clear that some
points are closer to the resonant manifold than others. This difference might be important
in numerics.

\begin{figure}[hbt]
\centerline{\epsfxsize=8.5cm \epsfbox{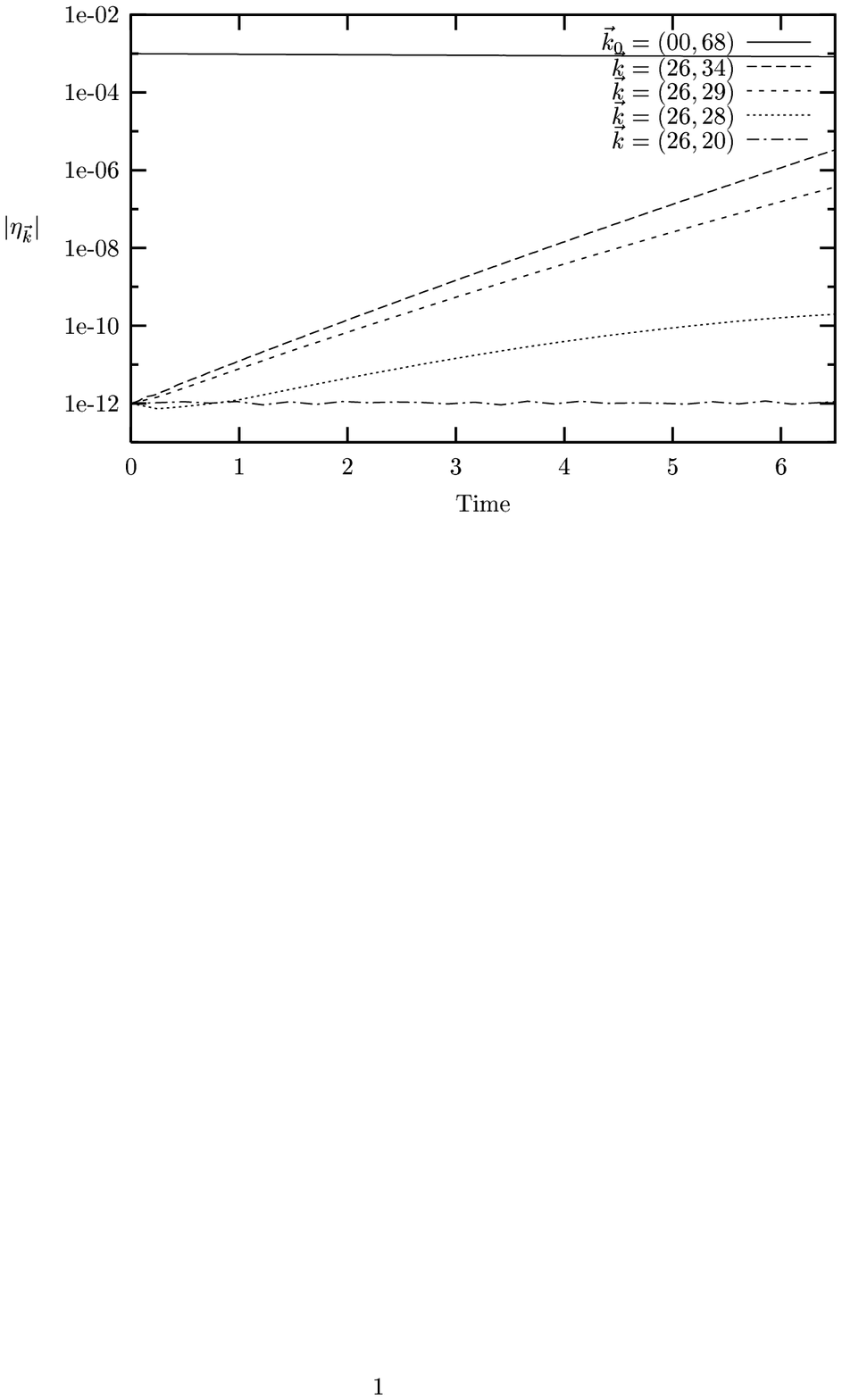}} 
\vspace{4mm}
\caption[]{\label{Growing}Fig.3. Evolution of various harmonics  for decaying wave $\vec k_0 = (00, 68)$.}
\end{figure}

\begin{figure}[hbt]
\centerline{\epsfxsize=8.5cm \epsfbox{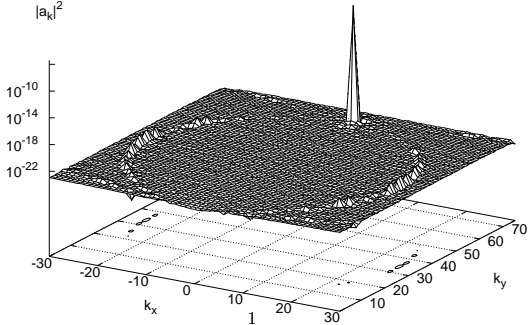}} 
\vspace{4mm}
\caption[]{\label{3d-1}Fig.4. Resonant harmonics starting to grow. Time t=1.4.}
\end{figure}

In the beginning, one can observe exponential growth of resonant harmonics in accordance with (\ref{exp-growth}) and (\ref{lambda}).
This is shown in Fig.\ref{Growing} and Fig.\ref{3d-1}. Here one can clearly see that some harmonics are in resonance
and others are not. 

\begin{figure}[hbt]
\centerline{\epsfxsize=8.5cm \epsfbox{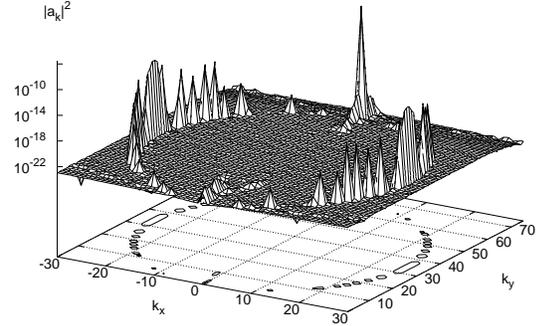}} 
\vspace{4mm}
\caption[]{\label{3d-2}Fig.5. Secondary decays start. Time t=11.}
\end{figure}

\begin{figure}[hbt]
\centerline{\epsfxsize=8.5cm \epsfbox{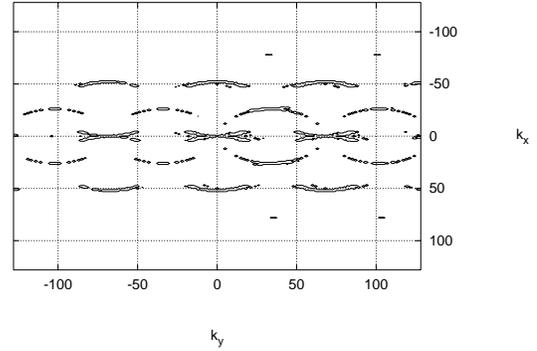}} 
\vspace{4mm}
\caption[]{\label{levels-1}Fig.6. The level lines for $|a_k|^2$. Secondary decays are
clearly seen. Time t=14.}
\end{figure}

\begin{figure}[hbt]
\centerline{\epsfxsize=8.5cm \epsfbox{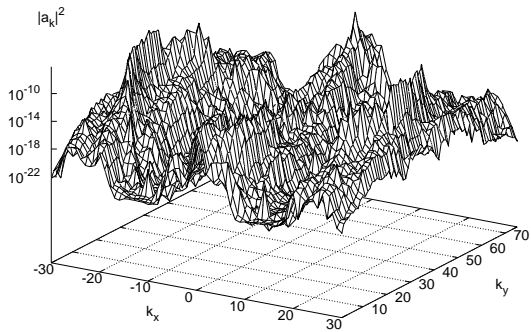}} 
\vspace{4mm}
\caption[]{\label{3d-3}Fig.7. Wave numbers spectrum at time t=57.}
\end{figure}

Than almost all harmonics in the resonant manifold become involved in the decay process (Fig.\ref{3d-2}).
Later, the harmonics that are the closest to the resonant manifold (compare with
Fig.\ref{resonance_curve_local}) reach the maximum level, while the
secondary decay process develops.
Waves amplitudes became significantly different. The largest amplitudes are for those waves with the maximal growth rate. One can see the regular structure generated by
the $\vec k_0$ wave in Fig.\ref{levels-1}.
After a while the whole $k$-space is filled by decaying waves, as shown in Fig.\ref{3d-3}.

Direct numerical simulation has demonstrated that the finite width of the resonance makes
discrete grid very similar to continuous. Of course, this is true only if the amplitude of the wave
is large enough, so that according to (\ref{lambda})
\begin{equation}
\label{condition}
\left|\frac{2\pi}{L_x L_y} M_{\vec k_1 \vec k_2}^{\vec k_0} A_{\vec k_0}\right| > \left|\frac{1}{2}\Omega_{k_1 k_2}^{k_0}\right|.
\end{equation}
As regards numerical simulation of the turbulence, namely, weak turbulence, the condition
(\ref{condition}) is very important. $A_{\vec k_0}$ has to be treated as the level of turbulence.

Authors thank Prof. E.A. Kuznetsov for very helpful discussions. This work was
supported by RFBR grant 03-01-00289, INTAS grant 00-292 ,the Programme
``Nonlinear dynamics and solitons'' from the RAS Presidium and ``Leading Scientific Schools of Russia" grant, also by US Army Corps of Engineers, RDT\&E Programm, Grant DACA 42-00-C0044 and
by NSF Grant NDMS0072803.

\end{document}